\documentclass{llncs}

\usepackage{amssymb}
\usepackage{graphicx}
\usepackage{enumitem}
\usepackage{multirow}
\usepackage{hhline}
\usepackage{url}
\usepackage{bbding} 
\usepackage{lscape}

\urldef{\mailsa}\path|lucas.gren@gu.se|
\urldef{\mailsb}\path|alex.wong@gu.se|
\urldef{\mailsc}\path|ejkredovisning@telia.com|
\newcommand{\keywords}[1]{\par\addvspace\baselineskip
\noindent\keywordname\enspace\ignorespaces#1}

\begin{document}

\mainmatter  

\title{Choosing agile or plan-driven enterprise resource planning (ERP) implementations\\  --- A study on 21 implementations from 20 companies 
}

\titlerunning{Choosing ERP Implementations}

\author{Lucas Gren\inst{1} \and Alexander Wong \inst{2}\and Erik Kristoffersson \inst{2}
}
\authorrunning{Lucas Gren, Alexander Wong, and Erik Kristoffersson}

\institute{Chalmers University of Technology and the University of Gothenburg \\
SE-412 96 Gothenburg, Sweden\\ 
\and
University of Gothenburg, School of Business, Economics, and Law.\\
 SE-405 30 Gothenburg, Sweden\\
\mailsa\\
\mailsb\\
\mailsc\\}

\toctitle{Lecture Notes in Computer Science}
\tocauthor{Authors' Instructions}
\maketitle

\begin{abstract}
Agile methods have gotten a good reputation for managing projects in many different sectors. A challenge among practitioners in the ERP (Enterprise Resource Planning) domain, is to decide if an agile method is suitable or not for new projects. This study investigates how decision-makers select between agile and plan-driven ERP implementation projects in relation to known factors from previous research. We selected projects that the decision-makers assessed as successful, but borderline, agile and plan-driven projects already implemented and let project managers fill out a survey consisting of key agile or plan-driven characteristics. We found that the assessment made by decision-makers did not differ on any aspects except higher executive buy-in for process change, and the prioritization of low cost in the agile projects. This study highlights the difficulty in selecting implementation strategy for a large part of the ERP implementations, and the data show that the decision-makers could not make such a decision at an early point in time without contextual knowledge.
\end{abstract}

\section{Introduction}
Plan-driven methods encourage detailed planning from the beginning of a project. Its most well known instantiation is the waterfall process. The waterfall process typically enables good performance when requirements are predictable, technologies are feasible, and plans are irrevocable \cite{royce}. However, when this is not the case, the most logical solution is to simply evolve the product as the client's needs change along the project development process. This shows the need for a method to be more flexible than a formal waterfall approach. Hence the shifting trend to the more iterative or incremental style of agile methods \cite{kerzner}.

Another reason to why firms adopt agile methods is the rapidly changing market situation of today. Agile methods offer a combination of disciplined execution and innovation. Irrespective if companies are selling products or consulting services, meeting the market's demand becomes increasingly important. The embedded incremental approach in agile methods makes it easier to meet market demand and adapt to its change. What was set as a requirement a year ago, can become obsolete when it reaches the market. When working with plan-driven project methods one face a risk that time and resources are invested in the wrong things. Agile methods reduce this risk \cite{holmstrom}. 

What makes agile methods interesting is that they seem successful. The IT industry is in general enthusiastic about agile methods and it is one of the most important trends. In academic literature agile methods are describes as ``examples of apparently major success stories'' \cite{agerfalk}. Qumer et al. \cite{qumer} claim that the phenomenon has meant ``unprecedented changes to the software engineering field''. A recent large empirical study from 2015 including 1002 projects shows that agility does contribute to project success (mainly efficiency and overall stakeholder satisfaction) \cite{serrador}. At the same time, Agerfalk et al. \cite{agerfalk} as well as Dingsoyr et al. \cite{dingso2012}, have emphasized that research yet has not answered the questions of how, why and in which contexts agile methods truly work. This creates challenges for managers who shall decide if to apply a plan-driven or agile project method. Research on agile methods is, in general, weak in theory \cite{dingso2012}, hence no dominant theoretical perspective on the understanding of agile methods has been established, such as when to apply an agile project method instead of a traditional plan-driven method, such as Waterfall. 

Most companies which claim to manage projects according to an agile method, are in practice applying a mixture of agile and plan-driven methods \cite{west}. This mixture is not necessarily bad as e.g.\ organizational context may be less adapted to a pure agile method. There are many factors which impact whether a plan-driven or an agile method is best suited for a certain project, but none of the ones found in the public domain has any empirical validation presented in connection to it. 

Agile development processes have spread to most aspects of software engineering and systems development. Also, at a larger company, such as SAP AG, the implementation of agile principles seem to increase customer satisfaction, give better results, and project also report effort savings of up to 20\%, according to an internal report \cite{sean}. The agile method says that smaller teams perform better and should not be more than five to nine members. Just like any big projects, SAP implementations need multiple small teams and interaction that fit the more classical standard SAP implementation milestones around the agile project \cite{sean}. This means that SAP implementations need more of a Water-Scrum-Fall process and keep their heritage of good architectural planing and rigorous testing. Many companies adapt the ``agile methods'' to their own context in such a way today \cite{west}.

At SAP AG they have experimented with using the agile method Scrum in the development phase of implementation projects since 2011. SAP divided the project types in three delivery models: 1.\ Assemble-to-Order Projects, 2.\ Design-Based Projects, and 3.\ Industrialized Projects, at the time of this research. The first one used an agile method as default due to the preferences of these types of project. The last one needed a rigorous planned approach and could leverage agile in some enhancements only. The tricky part was to know and navigate through the middle type of project, i.e.\ the Design-Based ones. The Design-Based projects could be helped by an agile method if the project had emerging requirements and were innovative. It would therefore be interesting to see if\slash decision-makers distinguish between projects that already have selected an agile method from projects that did not on a set of characteristics defined in previous research. This study put together agile discontinuing factors based on previous research (on agile potential \cite{datta}, discontinuing factors \cite{sidkyphd}, and agile\slash plan-driven risks \cite{boehm2003}) together with expert opinions from SAP practitioners involved in the projects, to see if the two types of project are assessed differently by the project managers on any of the items in the measurements. Since the decision of which method to use is at an early stage in the planning process (i.e.\ on a strategic level), the experts assessed which of the items (and added more if necessary to them) in the tool that could be assessed at the project stage they are in when they are forced make the decision.  

Hence, the aim of the research is to investigate if decision-makers can distinguish how agile and plan-driven projects differ in their strategic characteristics in the ERP domain by having the decision-makers assess the projects on a set of typical characteristics taken from previous research, and also let them assess the appropriateness of the selected implementation type. 

\paragraph{Research Question}
\begin{itemize}
\item ``Can decision-makers separate between how agile and plan-driven good-fit projects differ in their strategic characteristics in the ERP domain?
\end{itemize}

\section{Agile Project Management}\label{sec:apm}

\subsection{Agility -- The Strategic Perspective}\label{sec:backlitagile}
Even though the word agile is not present in the traditional management literature the underlying meaning of the concept is not new. The early organizational theorists, such as Taylor, Fayol, and Ford planned on the basis of a fundamental view that the world is predictable and therefore planning is essential \cite{eriksson}. After their great days of management in the early 1900s, the management literature moved increasingly towards the notion of greater flexibility. 

\cite{mintzberg} argues that strategic planning does not work and advocates strategic thinking for companies who want to be successful. The argument is that the outside world is not possible to predict, i.e., there is no single best way. Instead, the author claims that a strategy that is flexible and adaptable is beneficial. \cite{weick1993} have congruent ideas and specifies ``bricolage'' (do-it-yourself) skills as a valuable asset; the belief being that a successful employee should have the ability to act on his\slash her own, when necessary. In various articles and books Weick (e.g., \cite{weickbook,weick1993}) gives examples of people and situations in which this quality proved successful. \cite{agyris1977} promoted the idea of double-loop learning as a positive, but difficult to achieve, characteristic of an organization. By doing, evaluating, and doing again, an organization can constantly advance in the production and learn from the mistakes made. In the same context a successful Japanese management method in the 1980s was produced, called Kaizen, emphasizing continuous improvement. It was sprung from the successful Japanese automotive industry and had a high impact on later management ideas \cite{eriksson}.

Agile methods emphasize the importance of employees to devote themselves more towards the production and less to the documentation work. Daily meetings and frequent customer contacts replaces a deliberate strategy and careful planning. Congruent to the Kaizen or the double-loop learning methods, the agile method can take different paths through a project. There is an acceptance for changing the plan. Bricolage is valued and strong similarities with autonomous groups can be observed. What possibly is new with the agile ideas is that the product itself may evolve in several directions. Kaizen, autonomous groups and Total Quality Management (TQM) are much about developing methods of production and improve the quality, but do not focus (in the short time perspective) on that the end product itself can take many different forms. In the software industry, where the agile methods were first introduced, this is a fact today \cite{agilecobb}.

The implications for working in an agile way are that you wish to receive usable results quickly; the product should be functional in an early stage and then developed continuously in collaboration with the client. According to \cite{agilecobb} the agile principles are usable when the project has undefined demands, or when the client does not really know what she or he asks for. The principles are therefore less suitable when the opposite situation occurs; agreements with a firm frame that are extremely specified. \cite{kotter} write that the biggest challenge for today's companies is to satisfy the demands of flexibility in a rapidly changing environment. They claim that every company that manages to survive the start-up process is optimized for efficiency rather than strategic flexibility. They compare a company's management to an operative system (suitable to the agile theme) that is optimized for a ``day-to-day business'', but less adapted to the rapidly changing environment. The reason why these types of methods have become so popular in software engineering is probably because companies in that business sector have to be agile or else they will die out and be replaced. In other sectors in the corporate world organizations have not yet met the extreme conditions and environment that exists in software engineering, but they are getting there. That is probably why the ``agile'' thinking first surfaced in management, were fully adopted by the software industry, and now spread back to management under a different name.

\subsection{Plan-Driven vs.\ Agile Method}\label{sec:planagile}



An agile method description is typically designed as a set of recommended practical arrangements (e.g.\ roles, meeting procedures, documents and other tangible organizational arrangements) which claims to have a positive impact on the participant's engagement, flexibility, and productivity. In agile methods, the importance of the development team's autonomy is stressed. The practical design of roles, meeting procedures etc.\ should continuously be adapted to local needs, both within the organization and in to the specific project, even during the ongoing project. But without a solid explanation of how the agile arrangements impact the participants, every local adaptation may become a conjecture according to \cite{dybaa}. The adaptations are often based on fragmented experiences and intuition, and with some bad luck they become ineffective or even counteract their purpose. \cite{nerur} explain the differences between the two methods in Table~\ref{tradplan}.

\begin{table}
\renewcommand{\arraystretch}{1.3}
\caption{Traditional vs.\ Plan-Driven \cite{nerur}}
\label{tradplan}
\centering
\begin{tabular}{p{0.22\linewidth}||p{0.33\linewidth}||p{0.33\linewidth}}
\hline
\bfseries   & \bfseries Traditional & \bfseries Agile\\
\hline\hline
\bfseries Fundamental Assumptions & Systems are fully specifiable, predictable, and can be built through meticulous and extensive planning.  & High-quality, adaptive software can be developed by small teams using the principles of continuous design improvement and testing based on rapid feedback and change. \\
\hline
\bfseries Control & Process centric & People centric\\
\hline
\bfseries Management Style & Command-and-control & Leadership-and-collaboration\\
\hline
\bfseries Knowledge Management & Explicit & Tacit\\
\hline
\bfseries Role Assignment & Individual -- favors specialization & Self-organizing teams -- encourage role interchangeability\\
\hline

\bfseries Communication & Formal & Informal \\
\hline
\bfseries Customer's Role & Important & Critical\\
\hline
\bfseries Project Cycle & Guided by tasks or activities & Guided by product features\\
\hline
\bfseries Development Model & Life-cycle model (Waterfall, Spiral, or some variation) & The evolutionary delivery model\\
\hline
\bfseries Desired Organizational Form \slash Structure & Mechanistic (bureaucratic with high formalization) & Organic (flexible and participative encouraging cooperative social action)\\
\hline
\bfseries Technology & No restriction & Favors object-oriented technology\\
\hline
\end{tabular}
\end{table}

\section{Research on success factors for ERP implementations}\label{sec:planagile}
An enterprise resource planning (ERP) system is an information system that today is enterprise-wide. It is a software used to effectively plan and manage all resources of the organization. The initial use of IT in production was to plan and manage material and were therefore initially called material requirements planning (MRP) systems. The demand for integrated information systems and increased competitiveness made ERP vendors extend their software to more aspects of the enterprise and renamed their software to ERPs (enterprise resource planning). An even further demand to also cover supply chain management (SCM), supplier relationship management (SRM) and customer relationship management (CRM) lead to calling the systems ERPII systems \cite{koh2011drivers}. 

There has been some studies conducted to explain what factors that make an ERP implementation successful. Koh et al. \cite{koh2011drivers} conclude that the strategic partnership between the vendor and the customer is the key success factor for any enterprise-wide implementation. 

Bradley \cite{bradley2008management} investigated project success in relation to previously suggested factors in relation to ERP implementations. The factors that differentiated between successful and less successful projects were: (1) choosing the right full time project manager, (2) training of personnel, and (3) the presence of a champion. Unsupported differentiators were (1) the use of consultants, (2) the role of management to reduce user resistance, (3) the use of a steering committee to control the project, (4) integration of ERP planning with business planning, (5) reporting level of the project manager, (6) active participation of the CEO beyond project approvals, (7) resource allocation, and (8) occasional project review.

To summarize, the related work found is in relation to plan-driven project management and we found no rigorous academic studies on the agile approach to ERP projects, only smaller statements on the existing integration of agile ideas into ERP software (see e.g.\ \cite{nagpal2015comparative}).

\section{Agile Fit Tools}\label{sec:related_work}
The following section presents three different tools, that have been suggested by researchers, to evaluate the suitability of an agile approach on a strategic level. The reason for selecting these three tools was that they were the only ones found with a high-level approach to agility included. We looked for overall characteristics to assess the suitable project approach at an early point in time. 

\subsection{Agility Measurement Index}\label{sub:agility_measurement_index}
Datta \cite{datta} describes an Agility Measurement Index as an indicator for determining which method of Waterfall, Unified Software Development Process (UP), or eXtreme Programming (XP) should be used. Where Waterfall is plan-driven, UP is considered to be in the middle, and XP is an agile method. The author suggests five dimensions of a software development project that should be taken into account:

\begin{itemize}
\item Duration (D): How far ahead the deadline is.
\item Risk (R): The impact of the project deliverable when it is in use, and it malfunctions.
\item Novelty (N): If it is a brand new context for this type of product.
\item Effort (E): Time the customer is willing to put into the project.
\item Interaction (I): The amount of interaction between group members and customers during the project.
\end{itemize}

Each dimension is given a minimum $x$ score and a maximum $y$ score according to how complex the different aspects are in each organization. After this an actual score $a$ is set for the specific project (based on the range between $x$ and $y$). The Agility Measurement Index $\mathrm{AMI}$ is then defined as, 

\begin{equation}
\mathrm{AMI} = \frac{\sum a_{i}}{\sum y_{i}}
\end{equation}

A low score means that the project has short duration, low risk, etc.\ and thus a Waterfall approach is suggested \cite{datta}. 

\subsection{Using Risk to Balance Agile and Plan-Driven Methods}\label{sub:risk}
Another useful way of looking at these different types of projects is to look at the risks of using one or the other for a specific project. Their first step is to apply risk analysis to see if the project fits any of the agile or plan-driven home-grounds \cite{boehm20}. The dimensions used for this assessment are:

\begin{enumerate}
\item $Application$: Primary goals, Size, and Environment.
\item $Management$: Customer relations, Planning and control, and Communications.
\item $Technical$: Requirements, Development, and Tests.
\item $Personnel$: Customers, Developers, and Culture \cite{boehm20}. 
\end{enumerate}

Regarding aspect 1 the agile home-ground is rapid value and responding to change. The plan-driven home-ground is in that case predictability, stability, and high assurance. Agile projects should also have smaller teams compared to good plan-driven fits. The environment is also turbulent with a high change rate and more focused on the own project. Aspect 2 includes that the customer relations in agile is intimate and focused on prioritized increments while plan-driven has as-needed customer interactions and focus on contracts. The planning and control mechanisms are more qualitative in agile projects and more quantitative in plan-driven ones. Regarding communications, the agile project is seen to have tacit interpersonal knowledge while the plan-driven has explicit documented knowledge. The technical aspects of point 3 is that requirements are more informal rapidly changing user stories in agile and built on an extensive design with a foreseeable requirements evolution in plan-driven. The development of agile is short increments and under the assumption the re-factoring is cheap. The plan-driven home-ground has documented test plans and procedures, which agile has not. Aspect 4 states that agile projects has collocated customers that are dedicated and the developers are much more technically skilled and a culture of comfort, empowerment with many degrees of freedom. Pure plan-driven project thrive on order and have comfort and empowerment via framework of policies and procedures \cite{boehm20}.

\subsection{Agile Adoption Framework}\label{sub:agile_adoption_framework}
In order to define which agile methods an organization is ready to use, Sidky \cite{sidkyphd} suggests a method he calls the Agile Adoption Framework. He motivates its use by arguing that even though there are many success stories in agile development, they are not really generalizable. Sidky \cite{sidkyphd} then criticizes the framework created by \cite{boehm20} presented above, since it addresses agility in its generic form and not the actual practices. The transition to agile principles is tricky on a number of aspects, according to Sidky \cite{sidkyphd}. These are to:

\begin{enumerate}
\item Introduce structure in a complex and unpredictable process like that of agile development.
\item Measure and assess agility independent of agile methods.
\item Accommodate project and organizational characteristics influencing agile adoption efforts.
\item Ensure that the framework guides adoption effort in an efficient and effective manner.
\end{enumerate}

Sidky \cite{sidkyphd} then makes claims that his approach deals with all these points. The approach is based on a tool that has two parts. The first part is called the Agile Measurement Index (it is the same name as Datta \cite{datta} uses, but it is a different tool) and serves the purpose of being:

\begin{itemize}
\item A tool for measuring and assessing the agile potential of an organization independent of any particular agile method.
\item A scale for identifying the agile target level.
\item Helpful when organizing and grouping the agile practices in a structured manner based on essential agile qualities and business values.
\item Able to provide a hierarchy of measurable indicators used to determine the agility of an organization.
\end{itemize}

The second part is the use of the Agile Measurement Index through a four-stage process that will assess, firstly, if there are discontinuing factors, secondly, an assessment at project level, thirdly, an organizational readiness assessment, and lastly, a reconciliation phase. In this study we are interested in the high-level discontinuing factors. 

The discontinuing factors (or deal breakers or showstoppers) are critical overall aspects that need to be in place for agile software development to have a chance at working well. These are:

\begin{itemize}
\item Inappropriate Need for Agility
	\begin{itemize}
	\item Historical Project Schedules and Budgets
	\item Challenges with current software process
	\item Rate of Change of Project Requirements
	\item Time to Market Needed for Project
	\end{itemize}
\item Lack of Sufficient Funds
	\begin{itemize}
	\item Availability of Funds
	\end{itemize}
\item Absence of Executive Support
	\begin{itemize}
	\item Executive Management Buy-in
	\end{itemize}
\end{itemize}

The three tools suggested in this section were synthesized into the survey used in this study, which is described in more detail in the next section. 


\section{Method}\label{sec:meth}
To investigate if decision-makers assess agile or plan-driven projects differently, we used the Agile Fit Tools presented in the previous section to develop a survey to be filled out by project managers for existing implementations, i.e.\ we collected real project data from current and past agile and plan-driven implementation projects. We were looking at good agile fits or good traditional fits only, i.e.\ we were only looking for the projects that are ideal at each side of the agile\slash traditional plan-based spectrum.

\subsection{Subjects}\label{sec:ss}
\begin{figure}
\label{fig:pieind}
\includegraphics[scale=0.54]{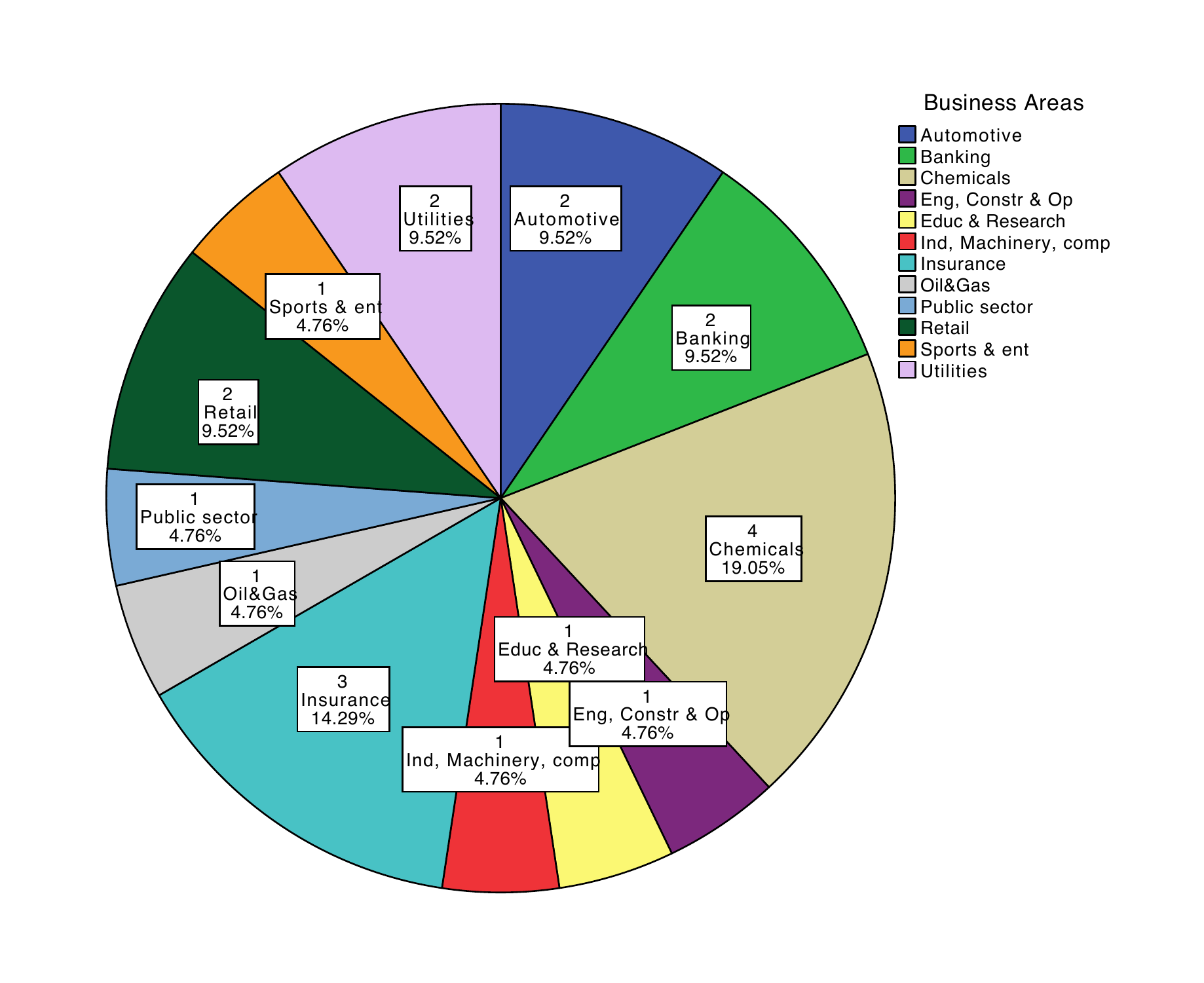}
\caption{Distribution of industries for the participating projects.}
\end{figure}

The research subjects, SAP customers, were selected by implementation experts within SAP. The selected companies were appointed by the regional agile implementation responsible from those continents, and the assessments were done by a person with an overview of the project on the SAP-side. The persons responsible for the continents were the ones giving feedback on the first version of the survey, and we received data from 6 organizations in North America, 3 in the Asia-Pacific region, 7 from Europe, and 5 from Latin America. The sample in this study was 21 projects from 20 companies (two project from the same company), and the 10 agile projects in this study were close to the only ones existing in the world within SAP (we know about a few more we did not get any response from). This means that the data in this study reflect almost the whole current population, and the industries represented in out data.  Figure~\ref{fig:pieind} shows the distribution of industries for the given sample. 


\subsection{Data Collection}\label{sec:dc}
The initial feedback from the persons responsible for each continent, except Western Asia and Africa, ($N=4$) were either via email or teleconference. The feedback regarded their experience in connection to what to alter or add\slash remove from the survey. The survey was then sent to the 21 managers involved in each project. Ten of them were stating that their projects were a good agile fit and eleven stated that their project was a good fit for the plan-driven approach. However, they were all design-based projects, which means they were all what was considered the middle-ground by SAP. The surveys were collected via an emailed offline spreadsheet. 

\subsection{Survey}\label{sec:survey}
When creating the survey we first summarized all categories included in the methods presented under the Agile Fit Tools section and the feedback received from the agile experts within SAP (the ``agile process'' responsible for each continent). We then compared them to each other and made sure all characteristics were covered in our own survey items. After this, the survey questions were tweaked to fit the SAP-specific context. The complete survey is shown in Table~\ref{fig:survey} shows the items as they are stated in related work and not the SAP-specific version. One critical difference between ERP implementations and many other projects is that they are always carried out at the customer-site. Therefore, critical questions about customer contact were left out due to the fact that these types of projects are always run within the customer organization.

The questions from Sidky \cite{sidkyphd} regarding Executive management buy-in were initially included but later removed from the discussion because all these items (items 17, 18, 19, 20, and 22 seen in Table~\ref{fig:survey}) included the words ``agile'' or ``agility''. It is evident that traditional projects get lower scores on items including this vocabulary and are therefore not very useful to analyze.


\subsection{Data Analysis}\label{sec:da}
Depending on the type of variable we conducted different analyses. Since we want to test if the survey items are different when it comes to chosen approach (agile or plan-driven) our dependent variable is categorical. Some of our independent variables were also categorical and with such a small sample size we also have the issue of not being able to assume normality, so we can not use Pearson's $\chi^2$ test for example. Therefore we used a Fisher's Exact Test or an Exact Contingency Table (i.e.\ randomization tests) for categorical variables and the Mann-Whitney U Test for the independent variables measured on an interval scale.

\section{Results}\label{sec:results}
The result of building the survey was to exclude some items due to the feedback received that such items can not be assessed at that strategic and early point in time. They were therefore left out and only the ones stated as possible to assess at an early stage were kept in the survey. Also, the questions regarding the respondents view of agility (items 17, 18, 19, 20, and 22) were expected to be significantly different since these items are about view of agility. These are important to have in the analysis criteria but are not presented here since they were obviously different for the collected samples (our statistical tests also confirmed this). 

The wording is slightly changed here to protect the intellectual property of the tool, but the meaning is the same however on a higher level of abstraction than the specific SAP case.

None of the Fisher's Exact Tests were significant at $\alpha = 0.05$ (two-tailed tests). 
We used the same method for the variable Priority of Time, Quality, and Cost. This categorical variable has six levels and we therefore used a $2\times6$ Exact Contingency Table (see Table~\ref{exact}). The sum of the probabilities of ``unusual'' tables was $p = 0.058$ so we conclude that it is unlikely that we observed such a table randomly, even if the value is slightly higher than 0.05, and we therefore reject the null hypothesis of independence (see e.g.\ \cite{everitt} for more details on this statistical method). To get an idea of the effect size, we can see that there is a 100\% chance that an agile project would prioritize Cost first (before both Quality and Time) for our sample. We can therefore reject the null hypothesis of Cost being of equal priority with 94.2\% probability in favor of the alternative hypothesis of that cost was more of a priority for the agile projects.

\begin{table}
\renewcommand{\arraystretch}{1.3}
\caption{Exact Contingency Table for item the prioritization of quality, time, and cost.}
\label{exact}
\centering
\begin{tabular}{c||c||c||c||c||c||c||c}
\hline
\bfseries   & \bfseries Q1T2C3 & \bfseries Q1T3C2 & \bfseries Q2T1C3 & \bfseries Q2T3C1 & \bfseries Q3T1C2 & \bfseries Q3T2C1 & Total\\
\hline\hline
Agile & 1 & 1 & 1 & 3 & 1 & 3 & 10\\
\hline
Non-Agile & 5 & 2 & 3 & 0 & 1 & 0 & 11\\
\hline
Total & 6 & 3 & 4 & 3 & 2 & 3 & 21\\
\hline
$p$ for agile & .17 & .33 & .25 & 1 & .5 & 1\\
\hline
\end{tabular}
\end{table}

Only one of the items in an interval scale were statistically significant, i.e.\ Insufficient current development process, meaning that the previously used project method was regarded as insufficient by the customers. The mean rank for the agile group was 12.79 and 7.41 in the plan-driven group ($p=0.035$).

We will now discuss the results and draw conclusions.

\section{Discussion}\label{sec:discussion}

This study set out to investigate if decision-makers can distinguish between good agile and plan-driven fit projects on a set of survey items in order to see if such decisions make sense at such an early point in time. 

The major finding is that we found little support in our data that the decision-makers can distinguish between agile and traditional projects the design-based category. We found a significant difference in that the projects had chosen an agile approach to cut costs. This is generally not a good idea since a change in process has a learning curve connected to it. An agile approach does not necessarily cut costs, but increases customer satisfaction and project success instead \cite{serrador}. Nevertheless, the participating agile projects were assessed as having costs as a higher priority then the plan-driven ones, a priority which can be explained by the different commercial setups between plan-driven and agile projects. While plan-based projects typically are fixed-price, and agile projects is more targeted to ``pay as you deliver,'' the need for cost control increases in agile projects.

A reason for trying a new implementation method could simply be to replace one that is not satisfactory. The question is if a change is inherently good just because it is a change, probably not always. The participating agile projects were assessed as having a higher degree of discontent with the current implementation method, meaning when the choice was made to implement the next one using the agile modifications, which, on the other hand, is a prerequisite for successful change, i.e.\ to believe the change is necessary \cite{lenberg2017initial}. 

This study therefore confirms the conclusion made by Koh et al. \cite{koh2011drivers} that it is very difficult for management to know, especially at an early stage, what will make an ERP implementation succeed or fail. The only significant differences we managed to find were prioritizing costs over time and quality, and if the customer see a need to change the current process of implementation. These are very general aspects of any change effort and we did not find that decision-makers could assess any real process-related differences between agile and plan-driven design-based projects. Therefore, we conclude that, in the ERP domain, we know which implementation strategy to choose in extreme cases (stable requirements or unknown and\slash or changing ones), but in the middle, it is very difficult to assess differences when the decisions were actually made.

Of course we have other success factors of an agile method, e.g.\ team capability, organizational culture, and empowerment of the team are important critical success factors for example \cite{chow,sheffield}. A collocated high performing team with good leadership would also most likely have a better chance at succeeding with an agile approach \cite{grenjss2,bradley2008management}. However, such information and in-depth analysis was not known by the decision maker in our study that had to select implementation method. Therefore, we conclude that the vendor needs deeper knowledge of the customer in order to select between the agile and plan-driven implementation strategies in many cases.

\subsection{Validity Threats}\label{sec:vt}
The largest threat against this study is the small sample size. Even if it is almost a full coverage of the intended population in the SAP context having data from other software implementations would, of course, be advantageous. However, we believe the interesting aspect in this case is the difference between agile and plan-driven projects as assessed by the decision-makers, not difference in software. From this perspective the given data sample is diverse with 21 different projects from four different continents of the world.

Another threat is that the assessments were conducted by the project responsible on the SAP-side. This means we only investigated the perception of the vendor, which might differ somewhat from the customers' perception of process and project success. There is also a possibility that the assessment made by the experts were not as informed as we hope. There might be other criteria in the organizations that made the project at hand perceived as a good fit for an agile or plan-driven approach. We also think that some items where we did not find significant differences should be more important for an agile approach, like the stability of the teams, or changing or unknown requirements for example. Perhaps, unstable teams and changing requirements cause as much trouble for both plan-driven and agile middle-ground projects in the ERP context since even the agile alternative should be seen as a hybrid. 

\section{Conclusions and Future Work}\label{sec:conclusions_and_future_work}
This paper set out to see if it is possible for decision-makers to assess the difference (on a strategic level) between good-fit middle-ground projects using an agile and plan-driven ERP implementation (i.e.\ no clear fit for either method) on a set of survey items based on previous research. Through a statistical analysis, we have found that these projects were not assessed as different on most strategic aspects. The only significant differences we found between these types of project were that they were assessed as having an insufficient current development process, and a prioritization of costs over time and quality. We conclude that it is relatively straight-forward which implementation strategy to choose in extreme cases (stable requirements or highly innovative projects), but in the middle, the decision-makers do not know at an early stage of an implementation project. These findings are important contributions to practitioners planning new projects as well as research by showing empirical data on the difficulty of knowing when to leverage agile implementations in the ERP domain. More studies are needed in other businesses to see how the agile and plan-driven middle ground projects might differ on other aspects than found in this study. We also need larger studies with both more in-depth investigation and larger samples (which will hopefully exist in the future).

We would also like to stress that we only looked at how agile and plan-driven projects perceived as successful in the organization differed. However, the usual definition of successful projects within SAP is in relation to the commonly used aspect of organizational impact and on time and on budget project completion \cite{bradley2008management}.

\section*{Acknowledgements}
This study was conducted with SAP AG (http://www.sap.com) and we would like to thank all the customers who shared information.

\bibliographystyle{splncs}
\bibliography{references}
\newpage
\begin{landscape}
\begin{table}
\centering
\caption{Survey used.}
\label{fig:survey}
\includegraphics[scale=0.45]{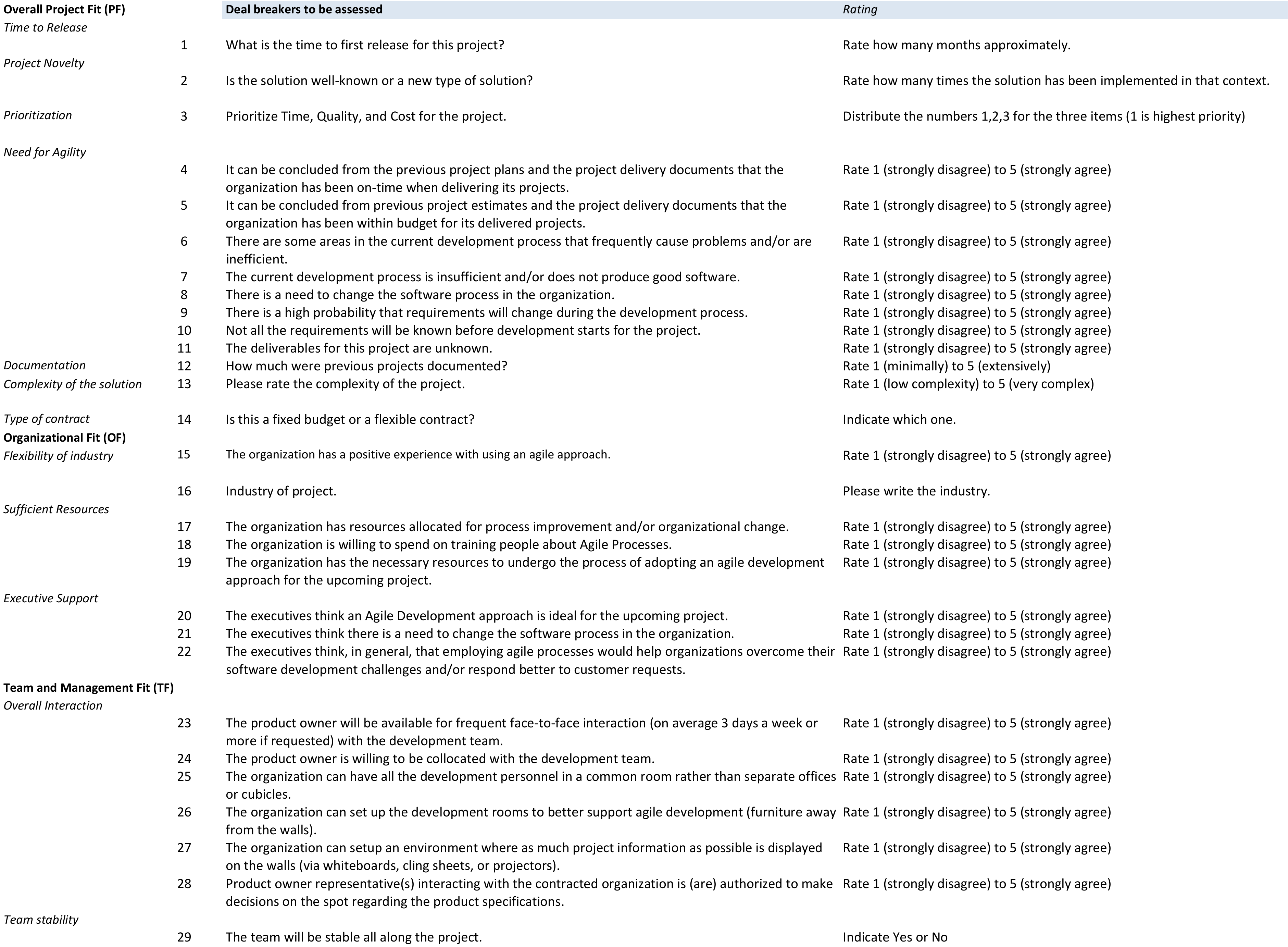}
\end{table}
\end{landscape}

\end{document}